\documentclass[%
 reprint,
 amsmath,amssymb,
 aps
]{revtex4-1}
\setcitestyle{super}

\usepackage{graphicx}
\usepackage[colorlinks=true, allcolors=blue]{hyperref}
\usepackage[export]{adjustbox}

\begin{document}

\title{Information-Optimal Sensing and Control in High-Intensity Laser Experiments}

\author{A.\ Döpp}
\email{a.doepp@lmu.de}
\affiliation{Ludwig-Maximilians-Universität München, Am Coulombwall 1, 85748 Garching, Germany}

\author{C.\ Eberle}
\affiliation{Ludwig-Maximilians-Universität München, Am Coulombwall 1, 85748 Garching, Germany}

\author{J.\ Esslinger}
\affiliation{Ludwig-Maximilians-Universität München, Am Coulombwall 1, 85748 Garching, Germany}

\author{S.\ Howard}
\affiliation{Ludwig-Maximilians-Universität München, Am Coulombwall 1, 85748 Garching, Germany}

\author{F.\ Irshad}
\affiliation{Ludwig-Maximilians-Universität München, Am Coulombwall 1, 85748 Garching, Germany}

\author{J.\ Schröder}
\affiliation{Ludwig-Maximilians-Universität München, Am Coulombwall 1, 85748 Garching, Germany}

\author{N.\ Weiße}
\affiliation{Ludwig-Maximilians-Universität München, Am Coulombwall 1, 85748 Garching, Germany}

\author{S.\ Karsch}
\affiliation{Ludwig-Maximilians-Universität München, Am Coulombwall 1, 85748 Garching, Germany}

\begin{abstract}
High-intensity laser systems present unique measurement and optimization challenges due to their high complexity, low repetition rates, and shot-to-shot variations. We discuss recent developments towards a unified framework based on information theory and Bayesian inference that addresses these challenges. Starting from fundamental constraints on the physical field structure, we recently demonstrated how to capture complete spatio-temporal information about individual petawatt laser pulses. Building on this foundation, we demonstrate how Bayesian frameworks can leverage temporal correlations between consecutive pulses to improve measurement precision. We then extend these concepts to active sensing strategies that adaptively select measurements to maximize information gain, exemplified through Bayesian autocorrelation spectroscopy. Finally, we show how these information-optimal measurement principles naturally extend to Bayesian optimization. This progression represents a paradigm shift where measurement devices transition from passive data collectors to active participants in complex experiments.
\end{abstract}

\keywords{information theory, Bayesian inference, laser diagnostics, optimization, uncertainty quantification}

\maketitle
\section{INTRODUCTION}

The characterization and control of high-power laser systems have become increasingly challenging as these systems grow in complexity and reach unprecedented intensities exceeding $10^{23}$ W/cm$^2$. Traditional measurement approaches, developed for stable, high-repetition-rate systems, often fail when applied to petawatt-class lasers that operate at low repetition rates with significant shot-to-shot variations\cite{alonso_spacetime_2024}. This challenge has motivated the development of new measurement paradigms that maximize information utilization, embracing machine learning \cite{dopp2023data} and, on a deeper level, information theory. 

Information theory was pioneered by Claude Shannon in the context of communication channels\cite{shannon1948mathematical}, but broader applicability stems from recognizing that information itself is fundamentally about distinguishing between possibilities. Any measurement in laser physics -- from detecting a single photon to characterizing a petawatt laser pulse -- represents the selection of one outcome from a space of alternatives, and the information content is determined by how improbable that particular outcome was among all possibilities. This sets the stage for probability theory: to quantify information, we must know the probability distribution over potential measurement results. This means that high uncertainty (a broad range of possible outcomes) corresponds to low information gained from a single measurement. Conversely, a precise measurement (narrowing down the possibilities significantly) gives us high information.

Its probabilistic foundation connects information theory directly to Bayes' theorem, which can then be seen as describing how information accumulates through successive measurements. In logarithmic form, Bayes' theorem becomes elegantly additive: 
\begin{widetext}
\begin{equation}
    \underbrace{{\log P(\text{parameters}|\text{data})}}_{\text{posterior knowledge}} = \underbrace{{\log P(\text{data}|\text{parameters})}}_{\text{measurement information}} + \underbrace{{\log P(\text{parameters})}}_{\text{prior information}} - \underbrace{{\log P(\text{data})}}_{\text{normalization}}
\end{equation}
\end{widetext}
This reveals that our posterior knowledge is simply the sum of our existing prior information content and the information delivered by the measurement, with the normalization ensuring proper probability scaling. Each measurement thus contributes information bits that directly update our knowledge state, transforming the traditional view of measurement uncertainty into a quantitative framework for optimal information processing \cite{Note1}.

When implemented correctly, these Bayesian and information-theoretic approaches help us get closer to "information optimality" – meaning we extract the maximum possible useful information from each laser shot, which is critical for experiments where data is sparse or expensive to obtain. This paper outlines our emerging framework to utilize such information-theoretic approaches: from passive laser diagnostics to active information gathering and system optimization. 

\section{SINGLE-SHOT VECTOR FIELD CHARACTERIZATION}

\begin{figure*}[t]
    \centering
    \includegraphics[width=1.2\columnwidth]{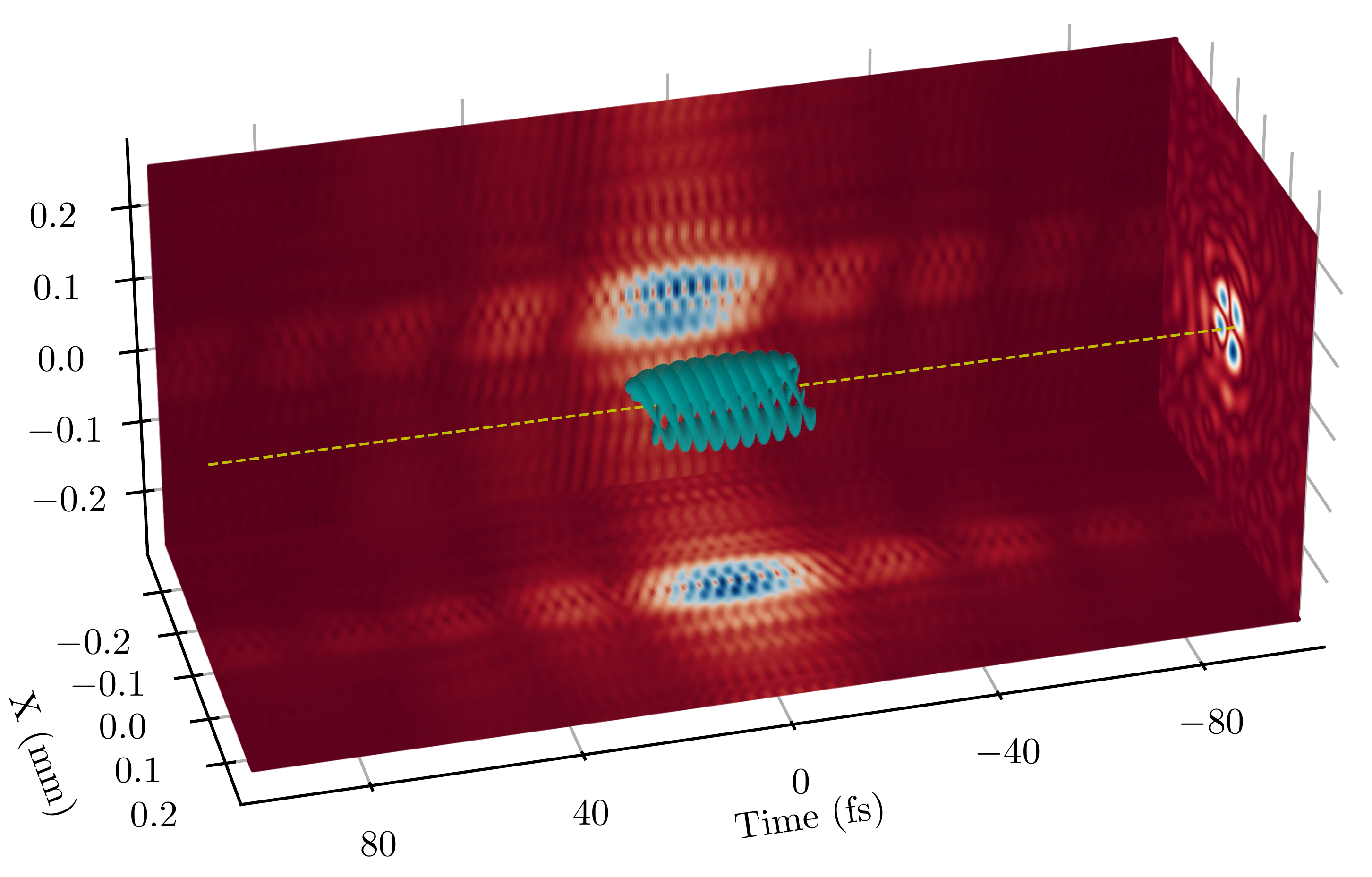}
    \caption{Electric field along $x$-direction in the spatio-temporal focus of a vortex beam as measured with RAVEN\cite{howard2025single}.}
    \label{fig:vortex}
\end{figure*}

The dilemma of modern diagnostics is aptly exemplified by the problem of measuring the electromagnetic vector field. In principle, this field forms a four-dimensional space-time manifold on which a six-dimensional "fiber" space of the electric and magnetic fields is embedded. In other words, spacetime is the "base" fabric, and at every single point in that fabric, there's an attached structure (called "fiber" -- not to be confused with optical fibers) that defines the local electric and magnetic field vectors. 

However, applying prior knowledge about the physics of these pulses makes the problem tractable. We first note that in a vacuum and for paraxial propagation, dimensionality reduces to the three-dimensional complex electric field, which itself decomposes into an intensity and phase term for each transverse position and color. A \textit{hyperspectral} measurement of intensity and phase, however, still poses a formidable challenge and common methods resort to extensive frequency or time domain scanning \cite{weisse2023measuring}.

The next, crucial insight enabling single-shot measurements comes from recognizing that ultra-intense laser pulses are subject to fundamental constraints in conjugate measurement domains. For petawatt-class lasers, the requirement for extreme intensity concentration at focus creates natural constraints in both spatial and temporal domains. Using principles from Fourier optics, we can determine the minimum resolution required in the near field to resolve a specified volume in the spatio-temporal focus. The Nyquist criterion directly relates these constraints: for a target focal volume covering $(-k_{x,max}, +k_{x,max}) \times (-k_{y,max}, +k_{y,max}) \times (-t_{max}, +t_{max})$, the required near-field resolution becomes $(\Delta x, \Delta y, \Delta \omega) = \pi\cdot (k^{-1}_{x,max}, k^{-1}_{y,max}, t^{-1}_{max})$. As both the bandwidth and aperture before focusing are well-constrained, the number of measurement points required is not only finite\cite{slepian_bandwidth_1976}, but also surprisingly small: In the spatial domain, approximately two measurements per Airy ring are required. The natural constraints imposed by the physics of intense light-matter interactions thus effectively reduce the dimensionality of the measurement problem.

Building on this insight, single-shot vector field measurements become possible through intelligent encoding of the complete electromagnetic field onto two-dimensional detectors. We recently demonstrated this for the first time using RAVEN (Real-time Acquisition of Vectorial Electromagnetic Near-fields)\cite{howard2025single}. Measurements are also performed on vortex beams (see Fig.\ref{fig:vortex}), where an overcomplete dictionary approach is used to identify the topological charge \cite{howard2025sparse}. In the spirit of probabilistic frameworks, RAVEN not only makes use of prior information in its design, but actively estimates the noise level of measurements to predict uncertainty of measurement variables.

\section{CONTEXTUAL INFORMATION IN SEQUENTIAL MEASUREMENTS}

\begin{figure*}[t]
    \centering
    \includegraphics[width=.9\columnwidth]{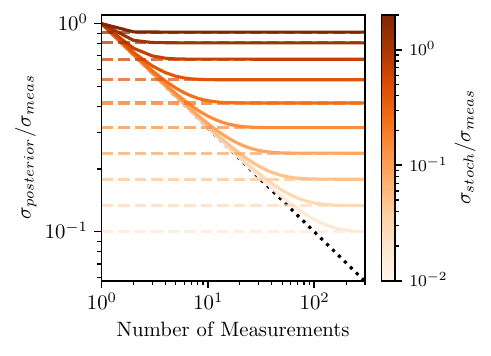}
    \includegraphics[width=.9\columnwidth]{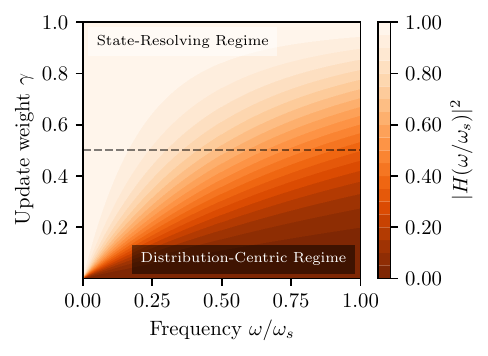}
    \caption{\textit{Left:} Evolution of the normalized posterior standard deviation ($\sigma_\text{posterior}/\sigma_\text{meas}$) over iterations for different ratios of process to measurement noise ($\sigma_\text{stoch}/\sigma_\text{meas}$, shown by color). Solid lines represent the Bayesian update process, while dashed lines indicate the corresponding asymptotic limits. \textit{Right:} Frequency response $|H(\omega/\omega_s)|^2$ as a function of normalized frequency $\omega/\omega_s$ and update weight $\gamma$. This plot illustrates how the system's response changes with different update weights, highlighting the trade-off between noise reduction and dynamic response in the Bayesian inference process. See Ref.\cite{esslinger2025bayesian} for more information.}
    \label{fig:magnitude_response}
\end{figure*}

While advanced diagnostics can capture a comprehensive snapshot of a laser pulse in a single instance, as discussed in the previous section, the information content of such a measurement is richer when considered within the context of a sequence. Laser systems evolve over time, and understanding the relationship between consecutive pulses provides both information about system stability and can increase the precision of measurements through the fusion of prior estimates and observations. An information-theoretic perspective allows us to quantify how knowledge of past states influences our understanding of current and future states\cite{esslinger2025bayesian}.

The state of a laser parameter $x_k$ at a given shot $k$ can often be viewed as a combination of a predictable component, which may depend on previous states, and an inherently unpredictable stochastic component $\varepsilon_k$ with variance $\sigma^2_\text{stoch}$. Each measurement $y_k$ of $x_k$ is further subject to measurement noise with variance $\sigma^2_\text{meas}$. Knowing the state of a previous pulse $x_{k-1}$ typically reduces our uncertainty about the current pulse $x_k$. This reduction in uncertainty is precisely the mutual information $I(x_k; x_{k-1})$. The more predictable the system, the higher this mutual information, and the less "new" information each subsequent shot provides about the underlying predictable evolution. The evolution of uncertainty over a series of measurements is bounded by the system's intrinsic stochasticity and the measurement precision, as illustrated in Fig. \ref{fig:magnitude_response} (left). This interplay highlights that true "single-shot" resolution depends not just on the device, but on whether a measurement can distinguish an individual pulse's state from the statistically expected behavior.

Beyond immediate temporal correlations, we can ask if older states, such as $x_{k-1}$, provide information about a future state $x_{k+1}$ even when the intermediate state $x_k$ is known. This is quantified by the conditional mutual information $I(x_{k+1}; x_{k-1} | x_k)$, sometimes referred to as second-order mutual information. If this quantity is non-zero, it implies that the system possesses a "memory" or trend that extends beyond simple shot-to-shot dependence. Such trends can be captured by models that incorporate more than just the last known state, for example, by estimating a local rate of change. The ability to resolve such trends depends on the measurement system's sensitivity to higher-frequency dynamics relative to its sampling rate, as depicted by the frequency response in Fig. \ref{fig:magnitude_response} (right).

This information-theoretic view emphasizes that the value of a measurement is contextual. It provides a quantitative basis for understanding how much information is truly gained from each shot, considering the system's inherent predictability and fluctuations. This understanding is a crucial prerequisite for designing more sophisticated, \textit{active} sensing strategies where the goal is to choose measurements that are maximally informative.

\section{INFORMATION-OPTIMAL MEASUREMENT SELECTION}

\begin{figure*}[t]
    \centering
    \adjincludegraphics[width=.8\linewidth,trim={{.\width} {.2\height} {.\width} {.17\height} },clip]{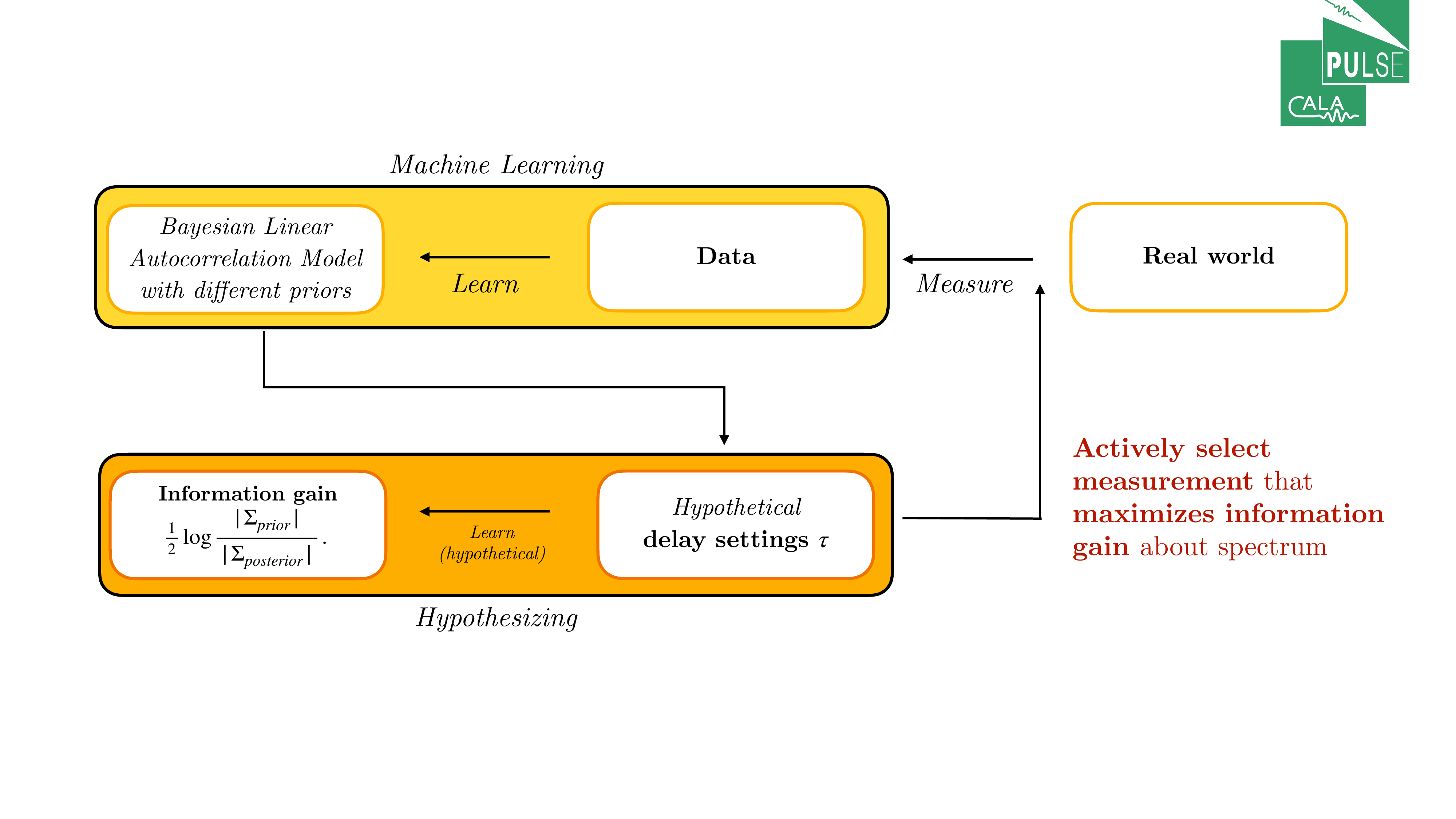}
    \caption{Sketch of the workflow of Bayesian autocorrelation spectroscopy.}
    \label{fig:bas}
\end{figure*}

The natural extension from studying sequential measurements is the quest for sets of measurements that provide maximal joint information\cite{schroeder2025information}. This represents a fundamental shift from fixed measurement protocols to dynamic, knowledge-driven approaches that optimize data collection in real-time.

The theoretical foundation rests again information theory, where measurement strategies are evaluated based on their ability to reduce uncertainty about quantities of interest. Traditional sampling theory, exemplified by Nyquist-Shannon sampling, emerges as a special case of this broader framework when operating with completely \textit{uninformed} priors. Uninformed essentially means we assume all possibilities are equally likely before we start measuring. However, if we possess some existing knowledge (e.g., the spectrum is likely centered around a known wavelength), this prior information can be exploited for more efficient measurement than standard Nyquist sampling would suggest. In Fourier transform spectroscopy (FTS), for example, conventional methods assume uniform uncertainty across all frequency components, leading to uniform sampling at the Nyquist rate. When prior knowledge about spectral structure is available, adaptive sampling can achieve equivalent reconstruction quality with far fewer measurements.

Bayesian autocorrelation spectroscopy is a generalization of FTS that allows for arbitrary prior distribution and uses information gain estimates to select optimal measurement positions, see Fig.\ref{fig:bas}. While information gain can be expensive to compute in many Bayesian settings, these computations can be expressed in closed analytical form for linear Gaussian noise models. This permits us to implement the sophisticated mathematical workflow of BAS in real-time settings. Applications span from molecular fingerprinting in medical diagnostics to hyperspectral imaging and optical vector field characterization. In each case, the adaptive approach provides superior performance while maintaining rigorous uncertainty quantification throughout the measurement process. The framework's ability to seamlessly integrate multiple information sources enables novel hybrid measurement schemes that combine the strengths of different modalities.

\section{BAYESIAN OPTIMIZATION OF LASER-PLASMA ACCELERATION}

\begin{figure*}[t]
    \centering
    \adjincludegraphics[width=.8\linewidth,trim={{.\width} {.2\height} {.\width} {.17\height} },clip]{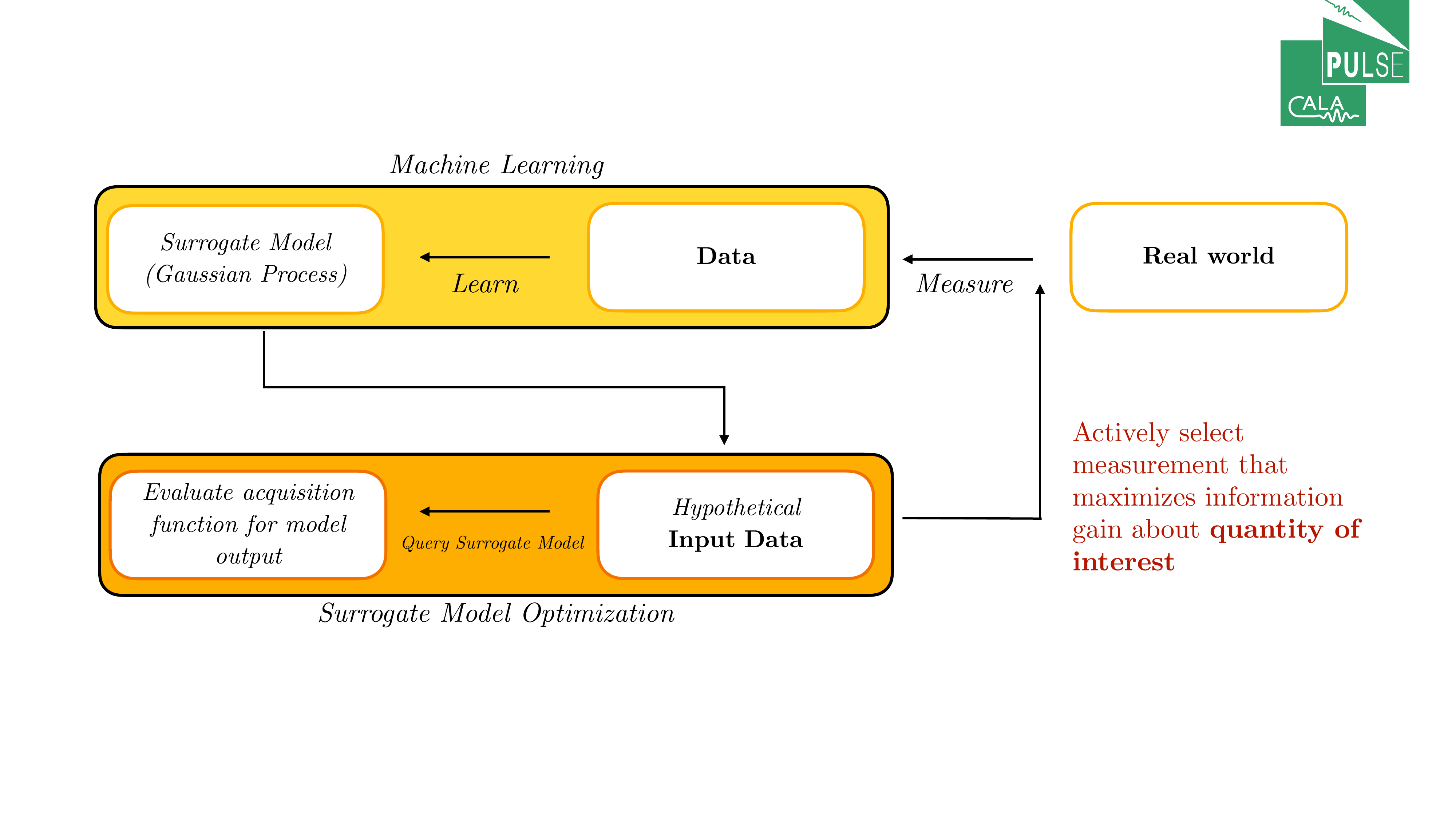}
    \caption{Sketch of the typical workflow used in Bayesian optimization.}
    \label{fig:bo}
\end{figure*}

While the previous example dealt with the metric of \textit{global} information gain, there are many examples in which one might only be interested in specific features in a distribution such as value or location of a maximum, see Fig.\ref{fig:bo}. These metrics can be approximated via the expected improvement, a widely used selection criterion in Bayesian optimization. A generalization is the expected hypervolume improvement, which can be loosely understood as optimizing the joint information of the \textit{Pareto front} of multiple properties. The "Pareto front" represents the set of solutions where no single objective (e.g., beam energy) can be improved without degrading at least one other objective (e.g., energy spread). 

We have demonstrated the use of such multi-objective optimization in a laser-plasma accelerator, both in simulations\cite{irshad2023multi} and experiments\cite{irshad2024pareto}. The approach is based on dynamically-updated, probabilistic surrogate models of the accelerator, which were used to predict beam energy, energy spread, charge, and stability. The resulting Pareto front reveals different operating regimes of varying laser-to-beam efficiency. Beyond system optimization, Bayesian approaches enable sophisticated control strategies such as energy tuning across wide ranges and inverse optimization, where users specify desired beam parameters and the system determines required settings. These capabilities represent a fundamental shift from manual system operation to autonomous, intelligent control.

\section{UNIFIED INFORMATION-THEORETIC FRAMEWORK}

The preceding sections have outlined a progression in how information can be optimally acquired and utilized in the context of high-intensity laser experiments. This journey starts from fundamental principles of measurement design and advances towards sophisticated, goal-directed optimization, following a natural hierarchy:

\begin{itemize}
    \item First, we leverage the most general (physics) priors and an understanding of system constraints to design measurement devices that efficiently capture information from individual events, such as in single-shot vector field characterization (Section 2).
    \item Second, we utilize contextual information, particularly from consecutive observations, to refine our knowledge and distinguish between predictable system dynamics and stochastic fluctuations, thereby enhancing measurement precision and interpretation (Section 3).
    \item Next, this evolves into strategies where we actively select measurements to maximize the overall information gain about the entire system, as exemplified by information-optimal adaptive sensing techniques like Bayesian Autocorrelation Spectroscopy (Section 4).
    \item Finally, the focus shifts to concentrating on acquiring information that is specifically relevant to a user-defined application or optimization goal, leading to methods like Bayesian Optimization for system tuning (Section 5).
\end{itemize}

This entire progression -- from foundational measurement design to targeted information gathering for optimization -- can be understood and unified under a general information-theoretic framework that treats optimal interaction (encompassing both measurement and control) as a sequential decision-making process. The core idea is to choose actions that maximize the information gained about specific, task-relevant aspects, $\phi$, of the system under investigation, which is often assumed to be governed by an underlying generative model, $G(\theta)$.
An interaction with the system, denoted by an action $A_t$ (which could be setting a measurement parameter or a control input at time $t$), yields an outcome or observation $y_t$. The optimal strategy, or policy $\pi^*$, for choosing a sequence of actions $\{A_t\}$ is then one that maximizes the total mutual information $I$ between the task-relevant aspects $\phi$ and the sequence of observations $\{y_t\}$ obtained:
\begin{equation}
\label{eq:active_sensing_objective}
\pi^* = \underset{\pi}{\arg\max}\; I(\phi; \{y_t\}_{t=1}^T | \pi).
\end{equation}
This objective essentially guides the agent to interact with the system in such a way that it learns as much as possible about what specifically matters ($\phi$) for the task at hand.

This general framework naturally encompasses the specialized approaches discussed earlier. For instance, in Bayesian Autocorrelation Spectroscopy (BAS) (Section 4), the system model $G(\theta)$ relates to the underlying true spectrum $S(\omega)$, which is also the task-relevant aspect ($\phi = S(\omega)$). The action $A_t$ is the selection of a specific delay $\tau$ for the FTS measurement, and $y_t$ is the measured fluence. BAS explicitly chooses $\tau$ to maximize the information gain about $S(\omega)$, aligning directly with Equation \ref{eq:active_sensing_objective}. Similarly, Bayesian Optimization (BO) (Section 5) can be viewed through this lens. Here, $G(\theta)$ represents the true, often unknown, objective function that maps system input parameters to performance. The task-relevant aspect $\phi$ is related to the location of this function's optimum (e.g., the input parameters $X_{opt}$ that yield the best performance). The action $A_t$ is the choice of the next set of input parameters to test, and $y_t$ is the observed system performance. The acquisition functions used in BO, which guide the selection of $A_t$, serve as heuristics or direct implementations of strategies to maximize information gain about $\phi$, helping to efficiently reduce uncertainty about where the optimal settings lie. For example, entropy-based acquisition functions directly aim to reduce the uncertainty (entropy) over the location or value of the optimum\cite{wang2017max,irshad2024leveraging}.

This unification provides both theoretical insights and practical benefits. Theoretically, it reveals the deep connections between seemingly disparate measurement and optimization approaches, rooting them in the common principle of maximizing task-relevant information. It also clearly delineates between learning from existing data -- as done from theoretical physics to machine learning -- and the pursuit of novel information at the heart of experimental physics. The method thereby formalizes the experimenter's intuition into a mathematical framework to guide the development of novel, integrated strategies for measurement and control within coherent experimental frameworks. To conclude, the framework outlined here provides a foundation for future developments in intelligent experimental systems, autonomous optimization, and data-driven discovery in laser science and beyond. By treating measurement and control as information processing problems, we open new possibilities for advancing both fundamental understanding and practical applications of extreme light fields.

\acknowledgments

The authors acknowledge support from the German Research Foundation (DFG), the Federal Republic of Germany and Free State of Bavaria for CALA infrastructure funding, and the Max Planck Society. This work was supported by various projects including the Independent Junior Research Group "Characterization and control of high-intensity laser pulses for particle acceleration" and international collaboration funding. C.E. is funded by BMBF through the MACLIP project. J.E., F.I., and J.S. thank the International Max Planck Research School for Advanced Photon Science (IMPRS-APS) for support.

\end{document}